\documentclass[pra,twocolumn,amsmath,amssymb,floatfix,reprint]{revtex4-1}

\usepackage{graphicx}
\usepackage{dcolumn}
\usepackage{bm}
\usepackage[usenames,dvipsnames]{xcolor}
\usepackage{mciteplus}
\usepackage{braket}

\begin{document}

\title{Turbulence in rotating Bose-Einstein condensates}
\author{Julian Amette Estrada$^1$\email{julianamette@df.uba.ar},
  Marc E.~Brachet$^2$\email{brachet@physique.ens.fr} \&
  Pablo D.~Mininni$^1$\email{mininni@df.uba.ar}}
  
\affiliation{$^1$Departamento de F\'\i sica, Facultad de Ciencias
Exactas y Naturales, Universidad de Buenos Aires and IFIBA,
CONICET, Ciudad Universitaria, 1428 Buenos Aires, Argentina.\\
  $^2$ Laboratoire de Physique Statistique, \'{E}cole Normale
  Sup{\'e}rieure, PSL Research University; UPMC Univ Paris 06,
  Sorbonne Universit\'{e}s; Universit\'{e} Paris Diderot, Sorbonne
  Paris-Cit\'{e}; CNRS; 24 Rue Lhomond, 75005 Paris, France.}
\date{\today}

\begin{abstract}
Since the idea of quantum turbulence was first proposed by Feynman, and later realized in experiments of superfluid helium and Bose-Einstein condensates, much emphasis has been put in finding signatures that distinguish quantum turbulence from its classical counterpart. Here we show that quantum turbulence in rotating condensates is fundamentally different from the classical case. While rotating quantum turbulence develops a negative temperature state with self-organization of the kinetic energy in quantized vortices, it also displays an anisotropic dissipation mechanism and a different, non-Kolmogorovian, scaling of the energy at small scales. This scaling is compatible with Vinen turbulence and is also found in recent simulations of condensates with multicharged vortices. An elementary explanation for the scaling is presented in terms of disorder in the vortices positions.
\end{abstract}

\maketitle

\section{Introduction}

% Introduction to QT
Quantum turbulence corresponds to the chaotic and out-of-equilibrium dynamics of quantized vortices observed in Bose-Einstein condensates (BECs) and in superfluid helium. Turbulence in both physical systems was studied in laboratory experiments \cite{Coddington2003, Bewley2006, Henn2009, White2014, Tsatsos_2016}, as well as theoretically and numerically \cite{Nore1997, Lvov2010, Laurie_2010, ClarkdiLeoni2015, Shukla2019, Mller2020}.

% QT regimes
Under many circumstances, quantum turbulence is very similar to its classical counterpart, to the point that identifying their distinguishing features became a major research topic. Many times both display Kolmogorov scaling $E(k)\sim k^{-5/3}$ of the kinetic energy, even though the mechanism behind this scaling in the quantum regime is believed to be vortex reconnection at large scales and a cascade of Kelvin waves at small scales \cite{Lvov2010}, the latter mechanism being unavailable in classical turbulence. However, some experiments \cite{Vinen1957, Walmsley2008, Barenghi2014} show another regime known as Vinen turbulence (or ``ultraquantum" regime), with $E(k)\sim k^{-1}$ scaling and with no classical counterpart. In this regime a thermal counterflow is believed to play an important role in the dynamics. This scaling was also found in numerical simulations with counterflow \cite{Baggaley2012, Baggaley2012bis}, but more intriguingly, also more recently in simulations of BECs with an initial array of ordered vortices and no apparent counterflow \cite{Cidrim2017, Marino2021}, as well as in simulations of homogeneous superfluid turbulence \cite{Polanco_2021}.

% Rotation
Rotating BECs display many interesting regimes that connect the flow dynamics and steady states with condensed matter physics \cite{Fetter2008}, including ordered vortex lattices \cite{Fetter_2001, Cooper_2004} and global modes and waves which have no classical counterparts \cite{Tkachenko1965, Andereck1982, Sonin_2005}. In spite of this, or perhaps because of its complexity, turbulence in rotating BECs has not been thoroughly studied so far. A recent numerical study considered rotating turbulence in unitary Fermi gases \cite{Hossain2022}, finding differences in the dissipation mechanisms between fermionic and bosonic superfluids. But a detailed comparison against classical fluids is still lacking. In classical turbulence, rotation generates a significant change in the system dynamics. The flow becomes quasi-two-dimensional (2D), a steeper-than-Kolmogorov spectrum $E(k) \sim k^{-2}$ develops at small scales \cite{Waleffe_1993, Cambon_1997, Cambon_2004, Pouquet_2010} in which inertial waves play a central role, and at large scales the flow self-organizes in columns with an inverse cascade of energy \cite{Sen_2012, Clark_Di_Leoni_2020b}.

%Classical rotating turbulence
For a detailed discussion on the theory of classical rotating turbulence, see Ref.~\cite{BELLET2006}. In the limit of very rapidly rotating incompressible flows, and in infinite domains, the flow becomes strongly anisotropic and the energy is mostly contained in inertial waves. This allows for wave-turbulence descriptions of the system \cite{Galtier2003, BELLET2006}. In this rapidly rotating limit, the $\sim k_\perp^{-2}$ energy spectrum results only for wave vectors in Fourier space close to the plane perpendicular to the axis of rotation, and the energy is transferred solely from large to small scales. The inverse energy cascade (i.e., the self-similar preferential transfer of energy towards large-scale modes, and in particular, towards two-dimensional modes) vanishes in the limit or angular velocity $\Omega \to \infty$ and of the domain height $H \to \infty$ \cite{BELLET2006}. For moderate rotation rates $\Omega$, and for finite domain heights, the inverse energy cascade can be recovered (see \cite{Sen_2012, Clark_Di_Leoni_2020b}, and a rigorous wave turbulence study in \cite{Scott2014}). It is important to note that the latter is the regime of interest when comparing with rotating BECs, as condensates in experiments are constrained by an external potential, and as for very large values of $\Omega$ a quantum phase transition to a different many-body state that does not have a BEC is expected \cite{Fetter2008}.

% What we do
In this work we study turbulence in rotating BECs in the rotating frame of reference. We show that rotating quantum turbulence is fundamentally different from its classical counterpart. While it displays, as in the classical case at moderate rotation rates, an inverse cascade of energy at large scales, at small scales it displays an anistotropic emission of waves and an energy scaling compatible with the ultraquantum turbulence regime.

\section{Methods} 
\label{sec:methods}

\subsection{The rotating Gross-Pitaevskii equation}

We solve numerically the Gross-Pitaevskii equation (GPE) with a trapping potential $V({\bf r})$ in a rotating frame of reference. The rotating Gross-Pitaevskii equation (RGPE), which describes the evolution of a zero-temperature condensate of weakly interacting bosons of mass $m$ under this conditions, is
\begin{equation}
     i \hbar \frac{\partial \psi (\mathbf{r},t)}{\partial t} = \left[  -\frac{\hbar^2 \nabla^2}{2 m} + g |\psi (\mathbf{r},t)|^2 + V(\mathbf{r}) - \Omega J_z \right]\psi (\mathbf{r},t) ,
   \label{eq:RGPE}
\end{equation}
where $g$ is related to the scattering length, $\Omega$ is the rotation angular velocity along $z$, and $J_z$ is the angular momentum operator. This equation can be obtained from the usual GPE by applying the constant-speed time-dependent rotation operator $R(t,\Omega)$, and redefining the order parameter in the rotating frame as $\psi = R(t,\Omega) \psi'$, where $\psi'$ is the wave function in the non-rotating frame.

By means of the Madelung transformation \cite{Nore1997} this equation can be mapped to the Euler equation for an isentropic, compressible and irrotational fluid in the non-rotating frame of reference with an extra quantum pressure term. The transformation is given by
\begin{equation}
    \psi ' (\mathbf{r},t) = \sqrt{\rho(\mathbf{r},t)/m} \, e^{i S(\mathbf{r},t)}, 
\end{equation}
where $\rho(\mathbf{r},t)$ is the fluid mass density, and $S(\mathbf{r},t)$ is the phase of the order parameter, such that the fluid velocity in the non-rotating frame is $\mathbf{v} = (\hbar/m) \mathbf{\nabla} S (\mathbf{r},t)$. The resulting flow is thus irrotational except for topological defects where the vorticity is quantized so that $\oint_{\mathcal{C}} \mathbf{v} \cdot d \mathbf{l} = (2\pi \hbar / m) n$ with $n \in \mathbb{N}$, and where $\Gamma_0 = (2\pi \hbar / m)$ is the quantum of circulation. In the rotating frame, the velocity is given by $\mathbf{v}_R= (\hbar/m) \mathbf{\nabla} S (\mathbf{r},t)- \Omega \hat{z} \times {\mathbf r}$. Replacing this velocity in the Euler equation, or equivalently, applying a Madelung transformation to Eq.~(\ref{eq:RGPE}), results in the Euler equation for the fluid in the rotating frame, with the extra Coriolis and centrifugal forces (see also \cite{Sedrakian_2001}).

Note that while classical rotating turbulence is typically studied in incompressible regimes \cite{Cambon_1997, BELLET2006}, BECs are diluted gases and compressibility cannot be neglected. Nevertheless, the weakly compressible case (which goes beyond the cases considered in this study) could be of interest for the large-scale dynamics of superfluid helium.

\subsection{Waves in the non-rotating system}
\label{sec:waves_norot}

In the absence of rotation and for $V({\bf r})=0$, Eq.~\eqref{eq:RGPE} becomes the usual GPE. If this equation is linearized around an equilibrium with uniform mass density $\rho$, one finds the Bogoliuobov dispersion relation for sound waves
\begin{equation}
    \omega_B(k) = c k \sqrt{1 + (\xi k)^2/2},
\end{equation}
where $c = (g\rho/m)^{1/2}$ and $\xi = \hbar/(2 g m \rho)^{1/2}$ are respectively the uniform sound speed and coherence length \cite{pethick_smith}. 

In the presence of quantized vortices, using the Biot-Savart law one can find normal modes of the vortex deformation. These correspond to a set of helicoidal Kelvin waves with dispersion relation

\begin{equation}
   \omega_K(k_\parallel) = \frac{2 c \xi}{\sqrt{2} r_n^2} \left( 1 \pm \sqrt{1+ k_\parallel r_n\frac{K_0(k_\parallel r_n)}{K_1(k_\parallel r_n)}} \right),
    \label{eq: Kelvin dispersion relation }
\end{equation}
where $r_n$ is the vortex radius, $K_0$ and $K_1$ are modified Bessel functions, and $k_\parallel$ is the wave number along the direction of the vortex core. The radius $r_n$ can be estimated using theoretical arguments, or directly from the density profile in experiments or simulations and is $\approx 2\xi$ \cite{Nore1997, Andereck1982, ClarkdiLeoni2015}. Typically, the random orientation of quantized vortices in a BEC results in a dependence of Eq.~(\ref{eq: Kelvin dispersion relation }) on $k$ instead of $k_\parallel$. The presence of rotation will align vortices preferentially along $z$, making $k_\parallel = k_z$. Finally, note that this dispersion relation is the same as the classical one derived by Kelvin, but dependent on the quantum of circulation $\Gamma_0 = 2 \sqrt{2} \pi c \xi$ instead of on the circulation associated to the total flow vorticity.

\subsection{Waves in the rotating system}
\label{sec:waves_rot}

The presence of rotation modifies the system behavior. Above a threshold in $\Omega$, $\Omega_c = 5\hbar /(2 m R_\perp^2) \ln (R_\perp/\xi)$ (where $R_\perp$ is the condensate radius), the flow tries to mimic a solid body rotation \cite{Fetter2008}. As a result of the quantization, the flow can only accomplish this by generating a regular array of quantized vortices such that their total circulation equals that of the rotation. The array is known as the Abrikosov lattice, forcing the system into a 2D state. To obtain a solid-body-like rotation, the density of vortices per unit area must be $n_v = \Omega / (\sqrt{2}\pi c \xi)$. Tkachenko \cite{Tkachenko1965} found that for an infinite system ($V({\bf r})=0$) this lattice must be triangular to minimize the free energy. When perturbed, this lattice has normal modes called Tkachenko waves. For the triangular lattice the modes follow the dispersion relation 
\begin{equation}
    \omega_T^2 = \frac{2C_2}{\rho m} \frac{c^2 k^4}{\{ 4 \Omega^2 + [4(C_1 + C_2)/(\rho m)]k^2 \}},
\end{equation}
where $C_1$ is the compressional modulus and $C_2$ the shear modulus of the vortex lattice \cite{Baym2003}. There are two Thomas-Fermi limits for this expression: The so-called rigid limit corresponds to small $\Omega$ compared to the lowest compression frequency $c k_0$, where $k_0$ corresponds to the fundamental mode of the trap. The soft limit corresponds to $\Omega$ larger than $c k_0$, but smaller than $mc^2 / \hbar$. In this regime, the vortex radius is smaller than the intervortex distance, and compressibility cannot be neglected. This is the regime we consider in this study, whose dispersion relation can be approximated as ($\gamma \approx 4$) \cite{Baym2003}
\begin{equation}
\omega_T^\textrm{(s)} = \left[ \left( 1- \gamma \frac{\sqrt{2}\Omega \xi}{c} \right) \frac{\xi c^3}{8 \sqrt{2} \Omega }\right]^{1/2} k^2 .
\end{equation}

Both the Abrikosov lattice and Tkachenko waves were experimentally observed in previous studies, such as \cite{Coddington2003}. Although to the best of our knowledge there are no laboratory studies of rotating quantum turbulence in BECs, vortex lattices were observed as metastable states in numerical simulations of rotating classical turbulence in finite domains \cite{Clark_Di_Leoni_2020b}.

The Kelvin dispersion relation also suffers a modification in the presence of rotation. For a single quantized vortex in the rotating frame it becomes
\begin{equation}
    \omega_K^\textrm{(r)} = \Omega + \omega_K(k_\parallel), 
\end{equation}
For many vortices, the presence of the vortex lattice also affects this dispersion relation; expressions taking into account this effect can be found in \cite{Andereck1982}. 

\subsection{Energy, momentum, and vortex length}

From the energy functional that defines RGPE, the total energy can be decomposed as $E = E_{\rm k} + E_{\rm q} + E_{\rm p} + E_{\rm V} + E_{\rm rot} $, with kinetic energy $E_{\rm k} = \langle \rho v^2 \rangle/2$, quantum energy $E_{\rm q}= \hbar^2/(2m^2) \langle (\nabla \sqrt{\rho})^2 \rangle$, internal (or potential) energy $E_{\rm p}= g/(2m^2) \langle \rho^2 \rangle$, trap potential energy $ E_{\rm V} = \langle V \rho \rangle$, and rotation energy $E_{\rm rot} = - \Omega \langle \psi^* J_z \psi \rangle$. In all cases, the angle brackets denote volume average. Using the Helmholtz decomposition $(\sqrt \rho {\bf v})=(\sqrt \rho {\bf v})^{\rm (c)}+ (\sqrt \rho {\bf v})^{\rm (i)}$ \cite{Nore1997}, where the superindices c and i denote respectively the compressible and incompressible parts (i.e., such that $\nabla \cdot (\sqrt \rho {\bf v})^{\rm (i)}=0$), the kinetic energy can be further decomposed into the compressible $E_{\rm k}^{\rm c}$ and incompressible $E_{\rm k}^{\rm i}$ kinetic energy components. It is worth pointing out that this decomposition is used in classical compressible flows \cite{KidaOrszag1990}. For each energy, using Parseval's identity we can build spatial spectra and spatio-temporal spectra \cite{ClarkdiLeoni2015}.

Another quantity of interest is the incompressible momentum spectrum $P^\textrm{(i)}(k)$ \cite{Nore1997}. It has been seen empirically that in many flows and for sufficiently large wave numbers, $P^\textrm{(i)}(k)$ can be obtained from the momentum spectrum per vortex unit length of a single quantized vortex, $P_s^\textrm{(i)}(k)$, summing it as many times as the number of vortices in the system times their lengths \cite{Nore1997, Shukla2019}. Thus, the total vortex length $L_v$ can be estimated as
\begin{equation}
 \frac{L_v}{2\pi} = \frac{\int_{k_\textrm{min}}^{k_\textrm{max}} P^\textrm{(i)}(k)dk}{\int_{k_\textrm{min}}^{k_\textrm{max}} P_s^\textrm{(i)}(k)dk},
\end{equation}
where $k_\textrm{min}$ is a cutoff ($k_\textrm{min}=10$ in this study, as the contribution from smaller wave numbers is dominated by the trap geometry), and $k_\textrm{max}$ is the maximum resolved wave number. From $L_v$, the mean intervortex distance is $\ell = (\mathcal{V}/L_v)^{1/2}$, where $\mathcal{V}$ is the condensate volume. 

\subsection{Numerical simulations}

We solve Eq.~\eqref{eq:RGPE} under an axisymmetric potential $V({\bf r}) = m \omega_\perp^2 (x^2+y^2)/2 $, in a cubic domain with periodic boundary conditions along the rotation axis. The choice of the axisymmetric potential corresponds to the elongated limit of a cigar-shaped trap, and is chosen to limit the contamination of the trap geometry in the computation of axisymmetric turbulent quantities. We use a Fourier-based pseudo-spectral method with $N^3 = 512^3$ spatial grid points and the $2/3$ rule for dealiasing, and a fourth-order Runge-Kutta method to evolve the equations in time, using the parallel code GHOST which is publicly available \cite{Mininni2011}, in a cubic domain of size $[-\pi,\pi]L \times [-\pi,\pi]L \times [-\pi,\pi]L$ so that the edges have length $2\pi L$. To accomodate the non-periodic potential and angular momentum operator $J_z = x \partial_y - y \partial_x$ in the Fourier base in $x$ and $y$, we smoothly extend these functions to make them (and all their spatial derivatives) periodic \cite{Fontana_2020}, in a region far away from the trap center such that the gas density in that region is negligible. This also prevents the occurrence of Gibbs phenomenon near the domain boundaries. To do so, a convolution between the Fourier transform of $V({\bf r})$ or $J_z$ and a Gaussian filter in $k_x$ and $k_y$ is computed. The width of the filter was chosen empirically to minimize errors in $V({\bf r})$ and in $J_z$ in the region occupied by the condensate. In practice we used a width $\sigma = (N \Delta k)/17 $, where $\Delta k$ is the resolution in wave number space. With this choice, errors in the computation of $V({\bf r})$ and $J_z$ were almost constant and $\approx 10^{-7}$ in the region occupied by the condensate. Values of $\omega_\perp$ were also chosen to keep the condensate confined in the region of the $xy$ plane satisfying these errors. 

\begin{table}[]
\begin{ruledtabular}
\begin{tabular}{cccccc}
$\Omega \, [U/L]$ & $\Omega/\omega_\perp$ & $\Gamma_\Omega / \Gamma_0$ & $\Omega/\Omega_c$ & $\ell / R_\perp$ & Ro \\ \hline
$0$   & $0$    & $0$    & $0$   & $0.47$ & - \\
$0.6$ & $0.29$ & $12.6$ & $2.27$ & $0.55$ & $6.9\times 10^{-2}$ \\
$0.8$ & $0.37$ & $16.8$ & $3.03$ & $0.44$ & $5.6\times 10^{-2}$ \\
$1.0$ & $0.47$ & $23.0$ & $4.12$ & $0.31$ & $5.4\times 10^{-2}$ \\
$1.2$ & $0.60$ & $38.4$ & $6.63$ & $0.39$ & $7.5\times 10^{-2}$ \\
$1.3$ & $0.55$ & $27.8$ & $5.02$ & $0.48$ & $5.1\times 10^{-2}$ \\
$1.5$ & $0.56$ & $25.1$ & $4.66$ & $0.31$ & $4.5\times 10^{-2}$ \\
\end{tabular}
\end{ruledtabular}
\caption{Parameters of all simulations. $\Omega$ is the rotation angular velocity, $\Omega/\omega_\perp$ is the ratio of $\Omega$ to the frequency of the potential, $\Gamma_\Omega/\Gamma_0$ is the ratio of the circulation in $\Omega$ to the quantum of circulation, $\Omega/\Omega_c$ is the ratio of $\Omega$ to the critical value $\Omega_c$, $\ell/R_\perp$ is the ratio of the intervortex length to the condensate radius, and Ro is the Rossby number.}
\label{table:Simulations}
\end{table}

In the following we use dimensionless units. All parameters are obtained by fixing $c_0 = (g\rho_0/m)^{1/2} = 2 U$ and $\xi_0 = \hbar/(2 g m \rho_0)^{1/2} = 0.017 L$, both defined using the reference mass density in the center of the trap $\rho_0=1 M/L^3$. These quantities are scaled with a unitary length $L$, a mass $M$, and a typical speed $U$. Considering typical dimensional values in experiments with $L\approx 10^{-4}$ m and $c_0 \approx 2 \times 10^{-1}$ m/s \cite{White2014}, this results in $\xi_0 \approx 1.7 \times 10^{-6}$ m (for the dispersion relations shown below, the relations in Secs.~\ref{sec:waves_norot} and \ref{sec:waves_rot} are evaluated using mean values for $c$ and $\xi$ in the condensate, obtained from the mean mass density in the trap with $\Omega=0$, $\left<\rho({\bf r},t)\right>_{\Omega=0}$).

\begin{figure}
    \centering
    \includegraphics[width=8.5cm]{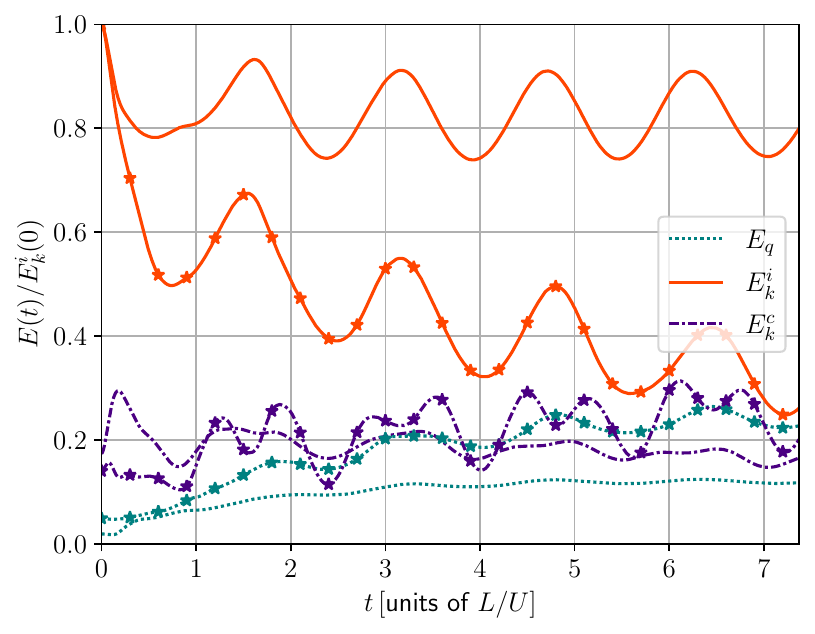}
    \caption{Time evolution of the compressible, incompressible, and quantum energy for two simulations (with $\Omega = 0$ for the lines with stars, and $\Omega = 1.2$ for the lines without markers). All energies are normalized by the initial ($t=0$) incompressible kinetic energy of each simulation.}
    \label{fig: Energy in time}
\end{figure}

It is important to note that we must prepare the system in a disordered initial state to have turbulence. Without such initial state, a non-rotating condensate should result in an equilibrium without quantized vortices, and a rotating condensate (with $\Omega > \Omega_c$) should result in an Abrikosov lattice. Moreover, none of these states can be readily accessed from the decay of the GPE or RGPE without proper initial conditions, as these equations have no dissipation (see, e.g., \cite{Verma_2022}). To obtain a turbulent state, we thus perturb an initial Gaussian density profile with a three-dimensional and random arrangement of vortices using the initial conditions described in \cite{Mller2020}, such that the kinetic energy spectrum peaks initially at $k\approx 5$ (i.e., $\approx 1/5$ of the domain size, leaving room is spectral space for self-organization processes). To reduce the emission of phonons, and to let the system decay into an initial condition compatible with the RGPE, we integrate this initial state to a steady state using a rotating real advective Landau-Ginzburg equation, which can be derived from Eq.~\eqref{eq:RGPE} following the method described in \cite{Nore1997} for the non-rotating case. The equation is
\begin{equation}
     \frac{\partial \psi}{\partial t} = \left[ \frac{\hbar \nabla^2}{2 m} - \frac{g}{\hbar} |\psi|^2 - \frac{V}{\hbar} + \frac{\Omega J_z}{\hbar} + \mu - i {\bf v} \cdot \boldsymbol{\nabla} - \frac{m|{\bf v}|^2}{2\hbar} \right] \psi ,
   \label{eq:RARGLE}
\end{equation}
where $\mu$ is the chemical potential and ${\bf v}$ the velocity field generated by the random arrangement of vortices. Note this equation corresponds just to the imaginary-time propagation of the RGPE, with a local Galilean transformation corresponding to the flow ${\bf v}$. The final state of this equation is then used as initial condition for RGPE. If we do not want a turbulent initial state (e.g., to get an Abrikosov lattice), we can integrate this equation with an initial Gaussian density profile and ${\bf v}=0$.

Table \ref{table:Simulations} lists the parameters of all simulations. As already mentioned the value of $\omega_\perp$ was varied with $\Omega$ to keep $R_\perp$ more or less the same. In all cases $\Omega/\omega_\perp \leq 0.6$, indicating the system is in or near a mean-field Thomas-Fermi regime \cite{Schweikhard_2004, Sonin_2005, Fetter2008}. Except when $\Omega=0$, $\Omega > \Omega_c$ (i.e., in the absence of turbulence the system displays a steady state with an Abrikosov lattice), and the circulation associated to the rotation $\Gamma_\Omega = \int \Omega dS$ is much larger than $\Gamma_0$. The intervortex distance is smaller than $R_\perp$ (the ratio $\ell/R_\perp$ is often accesible in experiments \cite{Coddington2003}), and a Rossby number defined as $\textrm{Ro} = u'/(2 \Omega R_\perp)$ (with $u'$ the r.m.s.~velocity in the rotating frame), which measures the inverse of the strength of rotation in classical turbulence, is small in all our rotating BECs.

\section{Results}

\begin{figure}
    \centering
    \includegraphics[width=8.5cm]{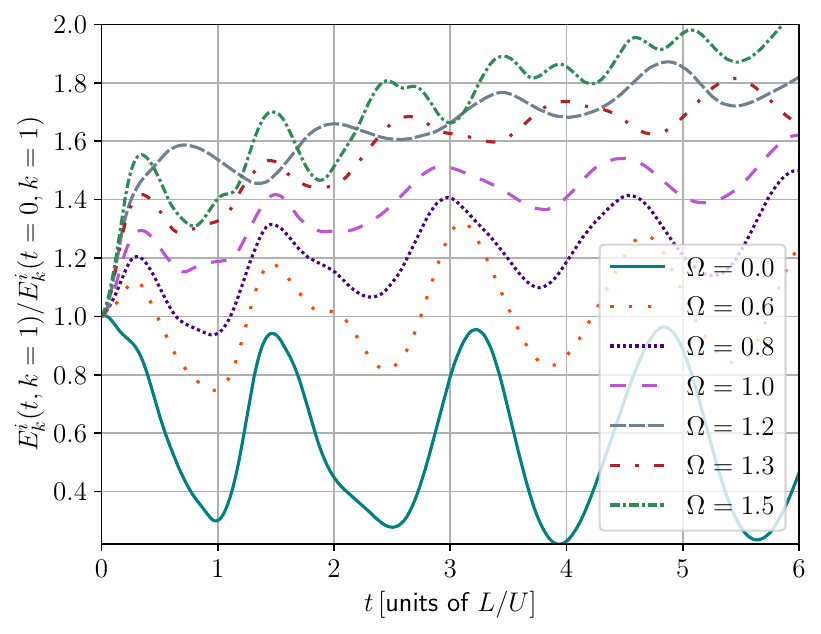}
    \caption{Incompressible kinetic energy at $k=1$ in all simulations, normalized by the initial ($t=0$) incompressible kinetic energy in the same Fourier shell.}
    \label{fig: energy evolution in k = 1}
\end{figure}

\subsection{The inverse energy cascade}
\label{sec: inverse cascade}

Figure \ref{fig: Energy in time} shows the time evolution of several energy components for the simulations with $\Omega=0$ and $\Omega=1.2$. All energy components display oscillations independently of $\Omega$, which are associated to a breathing mode of the condensate in the trap (indeed, we verified that this frequency is proportional to $2 \omega_\perp$, as expected for such mode \cite{Stringari1996}). Looking at the slow evolution, for $\Omega=0$ the incompressible kinetic energy decreases while the compressible and quantum energy increase. This is the result of the free decay of the turbulence: the incompressible kinetic energy is transferred towards smaller scales, and dissipated as sound waves. This results in the increase of energy in compressible motions, and in an increase of inhomogeneities which increase the quantum pressure. However, for $\Omega=1.2$ all energy components oscillate around a mean and approximately constant value, with a very small increase of the quantum energy at early times. This indicates less energy in the flow is being dissipated. Where is this energy going?

\begin{figure}
    \centering
    \includegraphics[width=8.5cm]{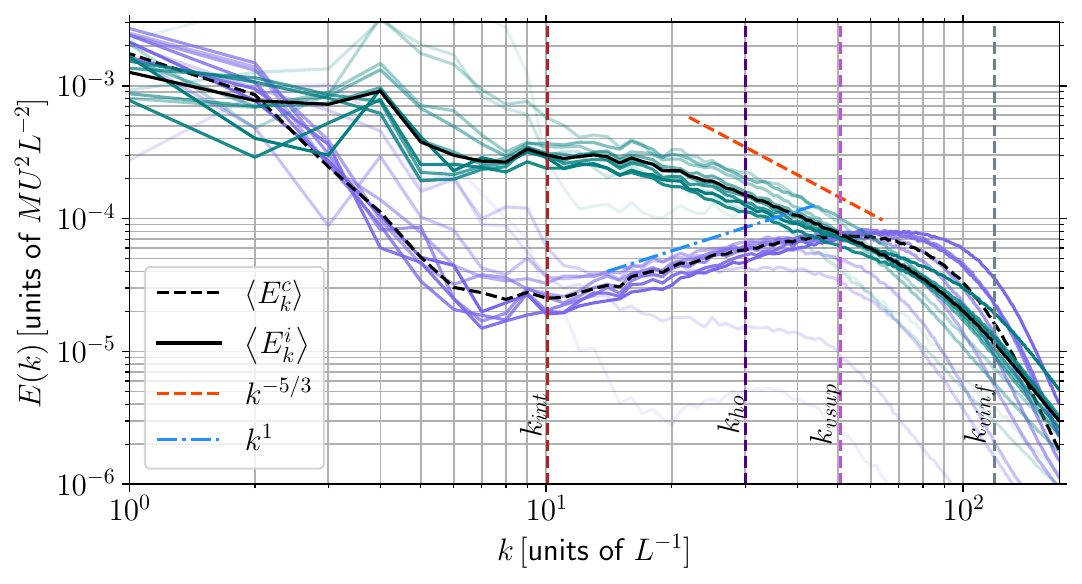}
    \includegraphics[width=8.5cm]{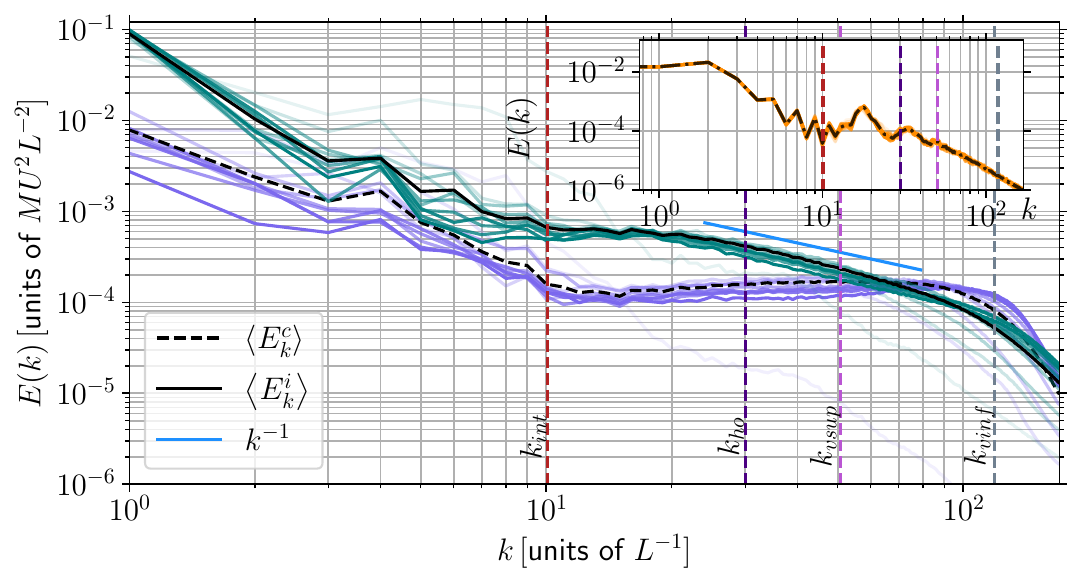}
    \caption{Incompressible (green, top curves at $k=10$ surrounding the time average with the solid black line) and compressible (purple, bottom curves at $k=10$ surrounding the time average with the dashed black line) kinetic energy spectra at different times (from light to dark as time evolves and time average over 2.5 breathing modes oscillations indicated with black lines) in a condensate with $\Omega=0$ (top), and with $\Omega=1.2$ (bottom). Several power laws are indicated as references by solid lines. The inset shows the incompressible kinetic energy spectrum of an Abrikosov lattice (i.e., a non-turbulent stationary solution). In all panels, vertical lines show characteristic wave numbers: $k_\textrm{int}$ associated to the intervortex distance, $k_\textrm{ho}$ associated to the condensate size for a non-interacting gas, and $k_\textrm{vinf}$ and $k_\textrm{vsup}$ associated to two measures of the intravortex scale.}
    \label{fig: spectrum}
\end{figure}

As shown in Fig.~\ref{fig: energy evolution in k = 1}, in the presence of rotation energy accumulates more and more at the largest available scale. The figure shows the time evolution of the incompressible kinetic energy at the gravest mode ($k=1)$ in all simulations. Leaving aside the oscillations, note that for $\Omega = 0$ energy in this mode decays slowly, while for $\Omega>0$, the stronger the rotation, the more the energy in this mode increases with time. In other words, the energy initially at $k\approx 5$ is transferred to the $k=1$ mode (i.e., to larger scales) instead of to larger wave numbers (smaller scales). As a result, less of the kinetic energy in the turbulent flow is available for dissipation as sound waves. This results from the quasi-two-dimensionalization of the flow in the presence of rotation, which results in an inverse energy cascade even in quantum turbulence \cite{Mller2020}, or, equivalently, in the condensation of the kinetic energy at the largest available scale in a process akin to Onsager's negative temperature states of an ideal gas of 2D point vortices \cite{Gauthier_2019, Johnstone_2019}. Thus, the first distinguishing feature of rotating quantum turbulence is its spontaneous evolution towards negative temperature states without the need for a change in the dimensionality of the trap.

The inverse energy cascade can be further confirmed in the spatial spectra in Fig.~\ref{fig: spectrum}, which shows the incompressible and compressible kinetic energy spectra at different times in the simulations with $\Omega=0$ and $\Omega=1.2$. While in the former case the incompressible spectrum peaks at all times at $k=4$, in the latter the same spectrum peaks at the smallest available wave number.

\begin{figure}
    \centering
    \includegraphics[width=8.5cm]{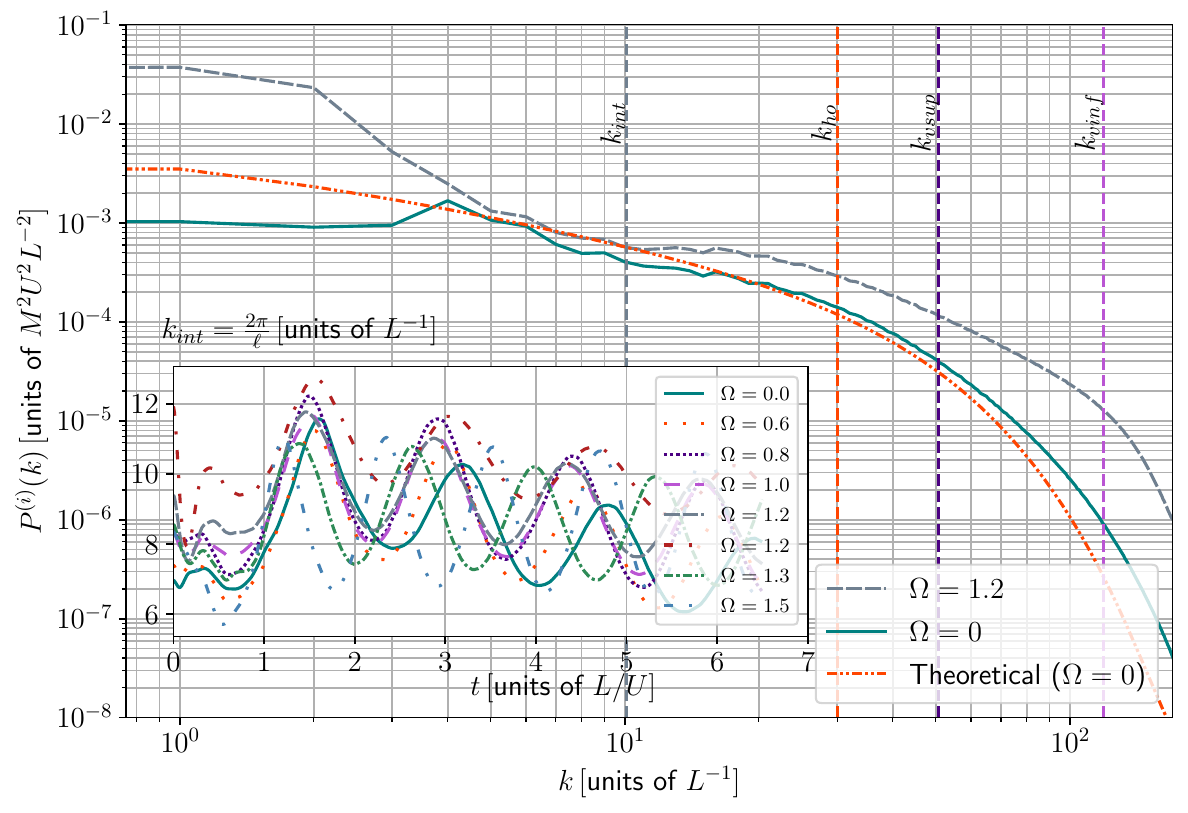}
    \caption{Momentum spectrum for $\Omega=0$ and $1.2$, compared with the theoretical momentum spectrum per unit length of one vortex \cite{Nore1997}, multiplied by the total vortex length in the simulation with $\Omega=0$. The inset shows the intervortex wave number as a function of time for all simulations.}
    \label{fig: momentum_spectrum_comparison}
\end{figure}

\subsection{The direct cascade subrange}
\label{sec: direct cascade}

For wave numbers $k>5$, the spectra in Fig.~\ref{fig: spectrum} display distinct power laws. When $\Omega=0$, the incompressible kinetic energy displays a range compatible with Kolmogorov $\sim k^{-5/3}$ scaling. The compressible kinetic energy displays a $\sim k^1$ scaling compatible with an axisymmetric (2D) thermalization, probably associated to the trap geometry. However, for $\Omega=1.2$ the spectra are very different. The incompressible direct cascade subrange is compatible with $\sim k^{-1}$ scaling, as in Vinen or ultraquantum turbulence. An inset in Fig.~\ref{fig: spectrum} also shows as a reference the incompressible kinetic spectrum of an Abrikosov lattice with $\Omega=1.2$ (i.e., of a non-turbulent stationary solution of RGPE), to show that its spectrum displays characteristic peaks and no clear $\sim k^{-1}$ scaling. The compressible kinetic spectrum in the rotating turbulent regime also changes its scaling and becomes flatter, as if the energy in sound modes reaches one-dimensional equipartition. As references, the figure also shows four characteristic, averaged in time, wave numbers: the intervortex wave number $k_\textrm{int}=2\pi/\ell$, the inverse harmonic trap length $k_\textrm{ho}$ for non-interacting bosons \cite{Dalfovo1999}, and $k_\textrm{vsup}$ and $k_\textrm{vinf}$, which correspond to the inverse lengths at which a single isolated vortex recovers respectively $0.9$ and $0.5$ of the mass density $\rho_0$ for $\Omega=0$. The direct cascade subranges take place for $k > k_\textrm{int}$ and $k< k_\textrm{vinf}$, and the direct $\sim k^{-1}$ scaling obtained with rotation is very different from the $\sim k^{-2}$ scaling observed in rotating classical turbulence \cite{Cambon_1997, Cambon_2004, Pouquet_2010}.

\begin{figure}
    \centering
    \includegraphics[width=8.7cm,trim=.2cm 0 0 0,clip]{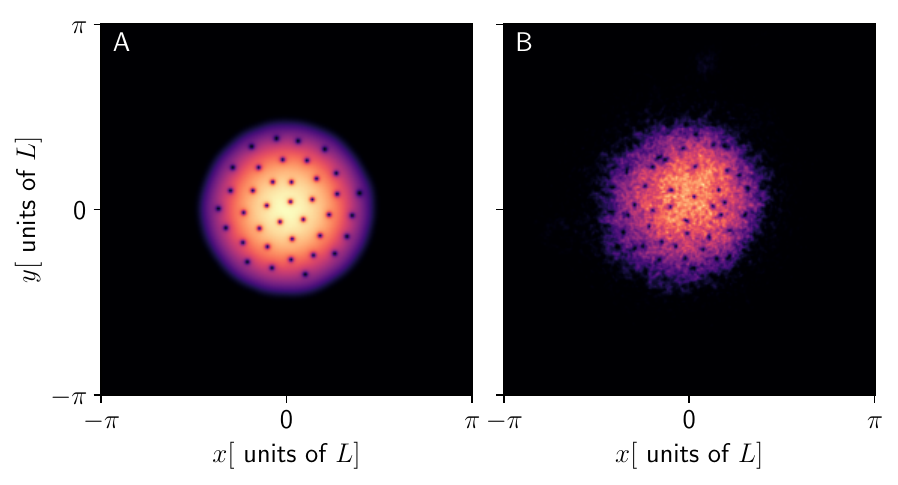}
    \includegraphics[width=7.5cm,trim=0 0 180 0,clip]{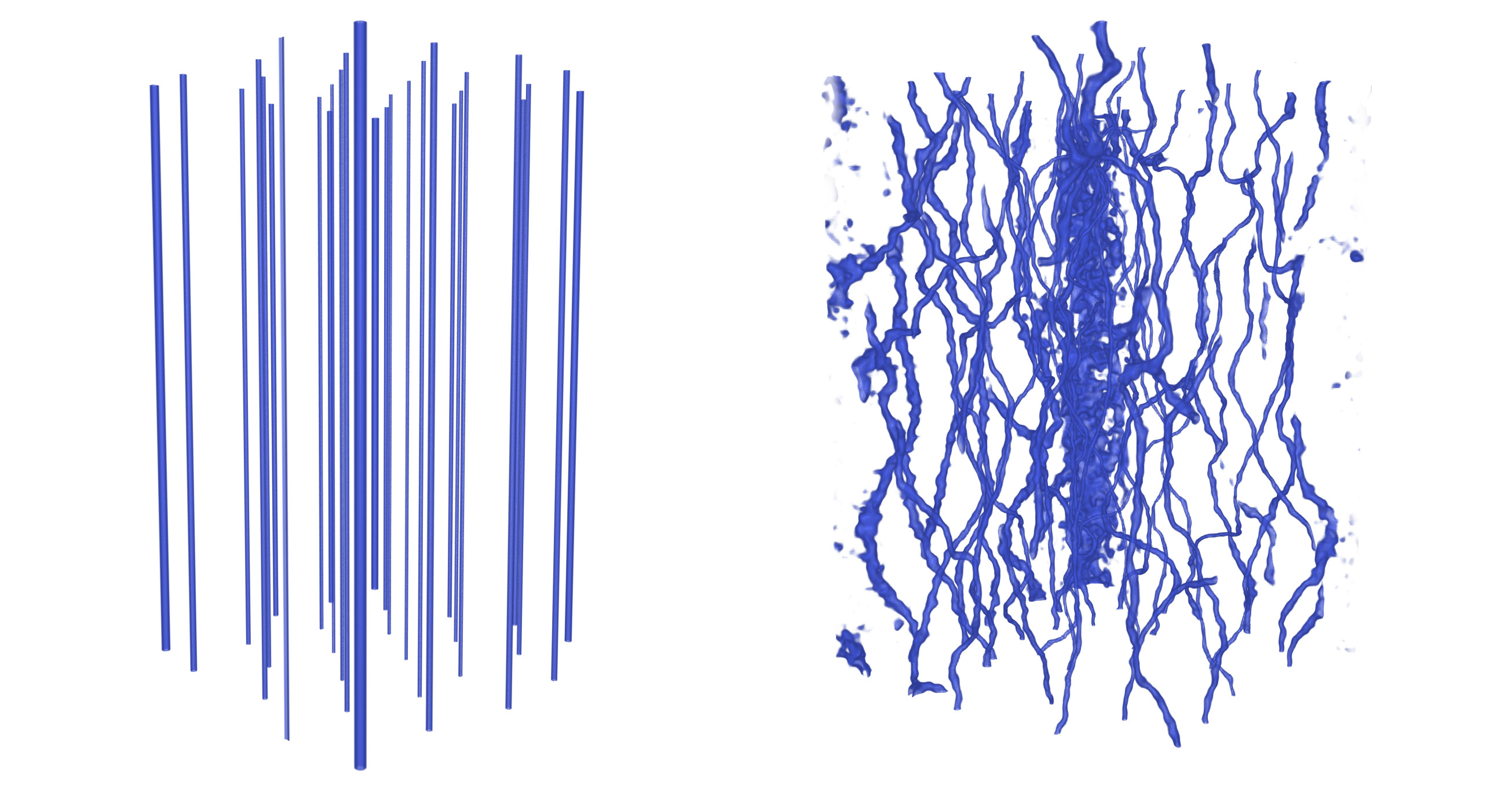}
    \caption{Density $\rho(x,y,z= \pi L)$ (top), and three-dimensional volume rendering of quantized vortices in the central region of the trap (bottom), for simulations with $\Omega = 1.2$ at late times in (A) a stationary regime, and (B) the turbulent regime. Volume renderings were done using the software VAPOR \cite{Clyne_2007}} 
    \label{fig: density profiles}
\end{figure}

A $k^{-1}$ scaling has been associated before to the presence of a counterflow \cite{Vinen1957}, to flux-less solutions \cite{Barenghi2016}, or to disorganized vortex tangles \cite{Barenghi_2014b, Polanco_2021}. In our case, the flux of energy towards small scales in the presence of rotation is substantially decreased, as evidenced by the accumulation of energy at large scales, and also by direct computation of the flux (not shown). Also, the vortex tangles in the flow in the presence of rotation change drastically. This is shown in Fig.~\ref{fig: momentum_spectrum_comparison}, which shows the spectrum of momentum $P^\textrm{(i)}(k)$ for $\Omega=0$ and $\Omega=1.2$, together with a theoretical estimation of the spectrum for a superposition of individual quantized vortices with the same total length (for $\Omega=0$). For $\Omega=0$ the shapes of the theoretical and observed mometum spectra are similar for $k \gtrsim 5$, but very different for $\Omega = 1.2$. This indicates that the vortex bundles indeed change in the presence of rotation. Differences at large scales (associated with the flow and trap geometry) can be expected in all cases; note in particular the excess of momentum at small wave numbers for $\Omega = 1.2$ which again confirm the large-scale self-organization. Differences at the smaller scales ($k>k_\textrm{ho}$) may be the result of contributions coming from the momentum field $\rho {\bf v}$ at the boundary of the condensed cloud. Indeed, in the presence of rotation there must be a net circulation generated by the vortex tangle in the condensate, which should be balanced with the circulation in a boundary layer. The inset in Fig.~\ref{fig: momentum_spectrum_comparison} shows the evolution of $k_\textrm{int}$ over time, calculated from the momentum spectrum. In all cases, on top of the breathing-mode oscillations, there is an initial increase of $k_\textrm{int}$ (and thus of $L_v$, the total vortex length) associated to vortex stretching.

\begin{figure}
    \centering
    \includegraphics[width=8.5cm]{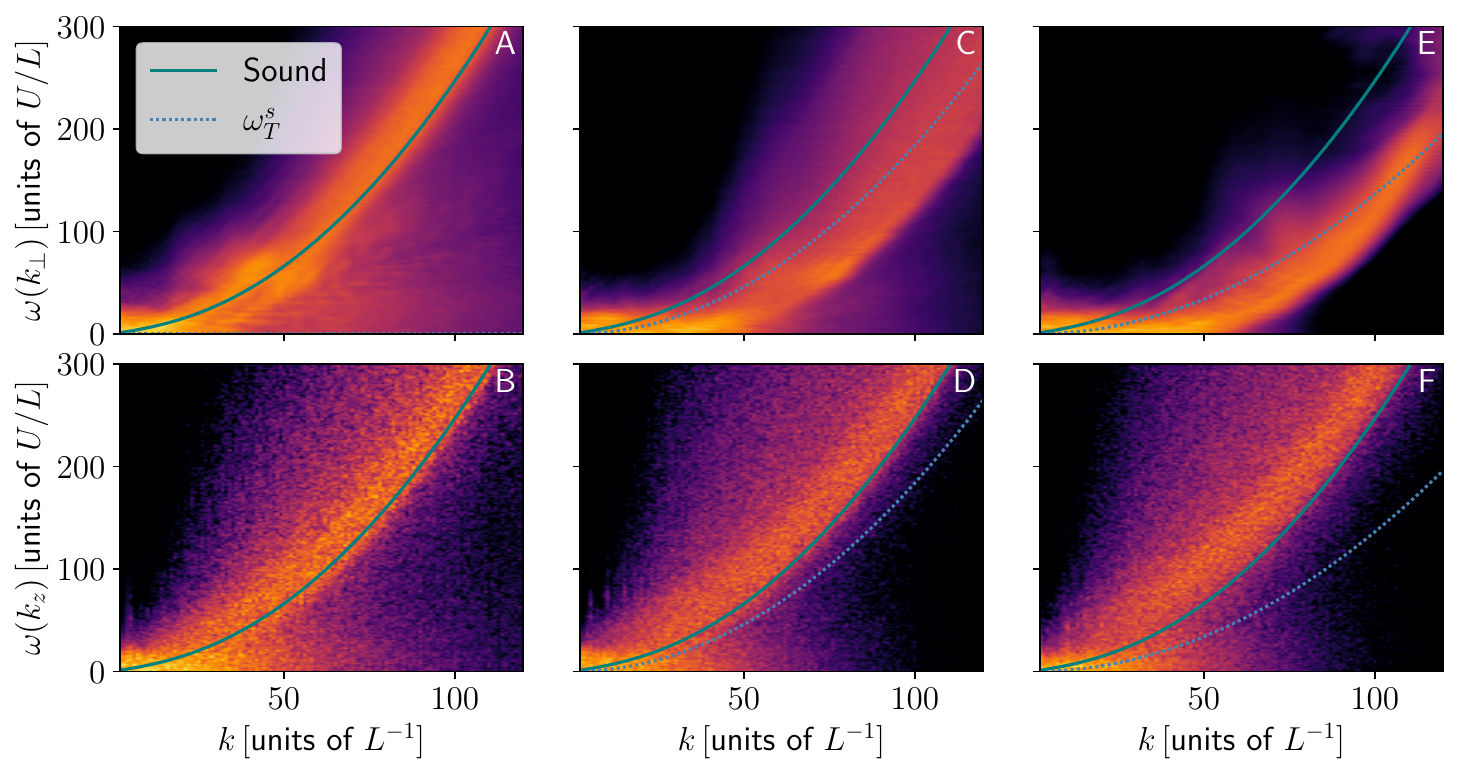}
    \caption{Spatio-temporal mass spectra for simulations with increasing $\Omega$, as a function of $k_\perp$ (for $k_z=0$) in A, C, and E, and as a function of $k_z$ (for $k_x=k_y=0$) in B, D, and F. From left to right, the columns show $\Omega = 0$, $1$, and $1.2$. As a reference the dispersion relation of sound and of soft Tkachenko waves are shown as references.} 
    \label{fig: spatiotemporal mass spectrum for different rotations}
\end{figure}

\begin{figure}
    \centering
    \includegraphics[width=8.5cm]{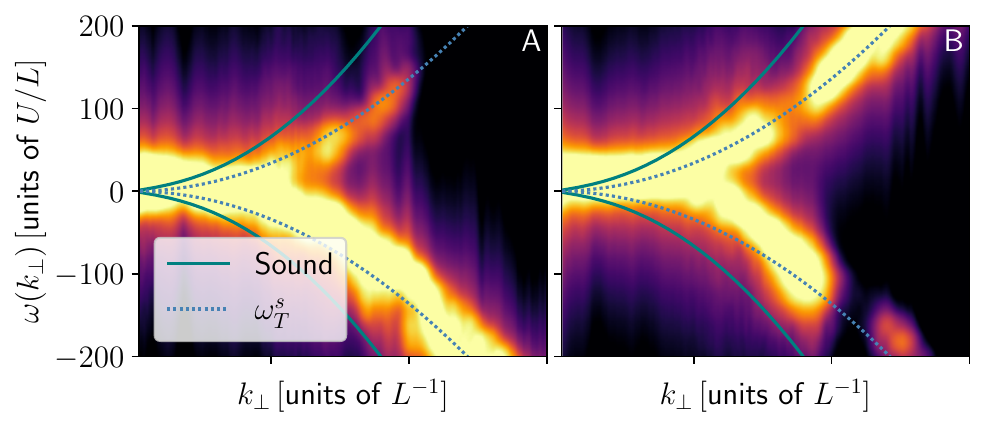}
    \caption{Spatio-temporal mass spectrum for $\Omega = 1.2$ as a function of $k_\perp$ (for $k_z=0$) centered at two different times. Note the pulsation between the positive and negative $\omega$ branches, as a mode moves outwards or inwards. As a reference, sound and soft Tkachenko dispersion relations are shown.} 
    \label{fig: spatiotemporal mass spectrum in time}
\end{figure}

However, and unlike homogeneous quantum turbulence, the $\sim k^{-1}$ scaling of the incompressible kinetic energy in the rotating case cannot be the result of unpolarized bundles of vortices (i.e., of randomly and independently oriented vortices \cite{Barenghi_2014b, Polanco_2021}). As explained before, the vortices in the rotating BEC must be polarized, and more or less aligned in order to approximate the solid body rotation. This is illustrated in Fig.~\ref{fig: density profiles}, which shows a horizontal cut of the mass density, and a 3D volume rendering of quantized vortices, for an Abrikosov lattice (i.e., in the non-turbulent stationary solution) and for the turbulent regime ($\Omega=1.2$). The latter system tries to mimic the former, with a quasi-2D bundle of vortices, albeit with disorder in the vortices' positions as well as with deformation in the $z$ direction (the axis of rotation). The $\sim k^{-1}$ scaling can thus be the result of the disorder in a quasi-2D system. Let's define ${\bf u}({\bf r}) = (\sqrt{\rho} {\bf v})^{(i)}$. The Fourier transform of the incompressible field generated by many quantum vortices can be written, using the translation operator, as $\hat{\bf u}(k) = \sum_j e^{-i {\bf k} \cdot {\bf r}_j} \hat{\bf u}_v(k)$, where $\hat{\bf u}_v(k)$ is the Fourier transform of the incompressible field generated by just one quantized vortex, and ${\bf r}_j$ the position of the $j$-th vortex. Then, the power spectrum of ${\bf u}$ is the angle average in Fourier space of
\begin{equation}
    \hat{\bf u}^*({\bf k}) \cdot \hat{\bf u}({\bf k}) = \sum_{ij} e^{-i {\bf k} \cdot ({\bf r}_i - {\bf r}_j)} |\hat{\bf u}_v({\bf k})|^2 ,
    \label{eq:Fourier}
\end{equation}
where the star denotes complex conjugate. If the vortices are organized in a lattice, the spectrum is dominated by the lattice spatial ordering (as in the inset in Fig.~\ref{fig: spectrum}). However, for a disorganized state with random positions, the sum in Eq.~(\ref{eq:Fourier}) reduces to the sum of the spectra of individual vortices, each with a $\sim k^{-1}$ scaling \cite{Nore1997, Polanco_2021}.

\begin{figure}
    \centering
    \includegraphics[width=8.5cm]{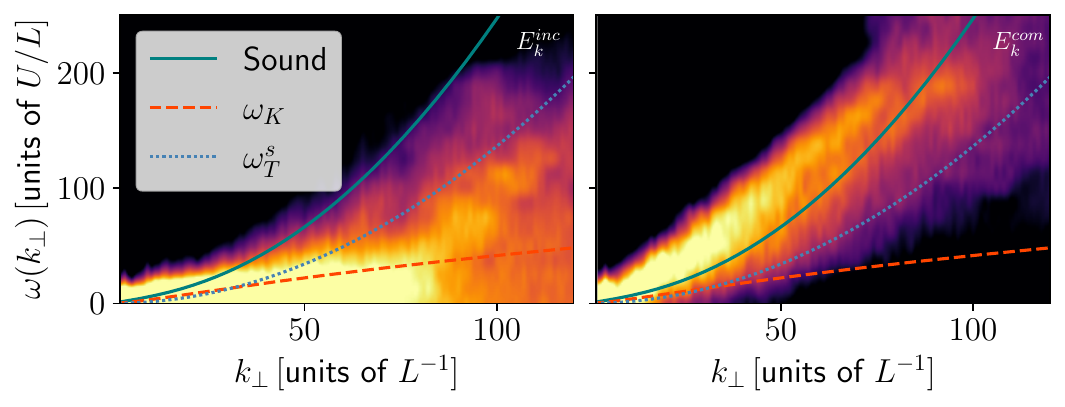}
    \caption{Spatio-temporal spectrum of the incompressible kinetic energy, $E_k^\textrm{(i)}(k_\perp, \omega)$, and of the compressible energy, $E_k^\textrm{(c)}(k_\perp, \omega)$ (for $k_z=0$) in a condensate with $\Omega = 1.2$. As a reference sound, soft Tkachenko, and Kelvin waves dispersion relations are shown. The solid-body mean rotation is removed from these spectra.} 
    \label{fig: Compressible and incompressible spectrum for omega = 1.2}
\end{figure}

\subsection{Wave emission}
\label{sec: emission of waves}

In non-rotating quantum turbulence, energy is transferred towards smaller scales through vortex reconnection and a Kelvin wave cascade \cite{Lvov2010}, and is finally dissipated through sound emission \cite{Kivotides_2001, ClarkdiLeoni2017}. The study of the waves excited by these flows can shed light on how energy is dissipated in the presence of rotation, and on the reasons for the different scaling laws observed in Fig.~\ref{fig: spectrum}. 

Figure \ref{fig: spatiotemporal mass spectrum for different rotations} shows the mass spatio-temporal spectrum \cite{ClarkdiLeoni2015} as a function of the frequency $\omega$, of $k_\perp=(k_x^2+k_y^2)^{1/2}$ (for $k_z=0$) or of $k_z$ (for $k_\perp=0$), for $\Omega=0$, 1, and $1.2$. Panels $A$ and $B$ show these spectra when $\Omega = 0$. Excitations accumulate near the dispersion relation of sound waves. When $\Omega$ increases, emission of waves changes drastically. In $k_z$, excitations still accumulate around sound waves: turbulence dissipates energy by emmiting sound in the $z$ direction. But in $k_\perp$ the dispersion relation shifts towards larger values of $k_\perp$ as $\Omega$ increases (panels C and E), and become closer to soft Tkachenko waves.

These modes in $k_\perp$ are not stationary. Figure \ref{fig: spatiotemporal mass spectrum in time} shows two spatio-temporal mass spectra as a function of $k_\perp$ in the simulation with $\Omega = 1.2$, for both positive and negative frequencies. Note the pulsation between positive and negative $\omega(k_\perp)$ branches as time evolves. In other words, modes are respectively of the form $\exp [i({\bf k}_\perp  \cdot {\bf r}_\perp - \omega t)]$ and $\exp [i({\bf k}_\perp  \cdot {\bf r}_\perp + \omega t)]$, or equivalently, the modes collectively propagate outwards or inwards. Interestingly, the alternation of energy between the positive and negative $\omega(k_\perp)$ branches is not visible in the simulation with $\Omega=0$. Thus, it must represents a global deformation of the vortex lattice on top of which turbulence develops (and also feeds with energy), the breathing mode possibly being part of it, and which can give a mechanism for energy dissipation as vortices move through this pulsation (i.e., it could act as an effective counterflow). 

Waves not only manifest in the mass spatio-temporal spectrum. The spatio-temporal spectra of the incompressible and compressible kinetic energies as a function of $k_\perp$ (for $k_z=0$) are shown in Fig.~\ref{fig: Compressible and incompressible spectrum for omega = 1.2}, computed after turbulence is totally developed and over half a breathing mode period. The spectra are computed after removing the solid-body rotation. The incompressible energy shows excitations at lower frequencies, near the Kelvin and soft Tkachenko dispersion relations, and with excitations at frequencies close to the soft Tkachenko modes observed in the mass spectrum in Fig.~\ref{fig: spatiotemporal mass spectrum for different rotations}, suggesting these modes correspond in part to inward or outward incompressible deformations. In the compressible energy, excitations are approximately compatible with sound modes, and with some power in Tkachenko frequencies.

\begin{figure}
    \centering
    \includegraphics[width=8.5cm]{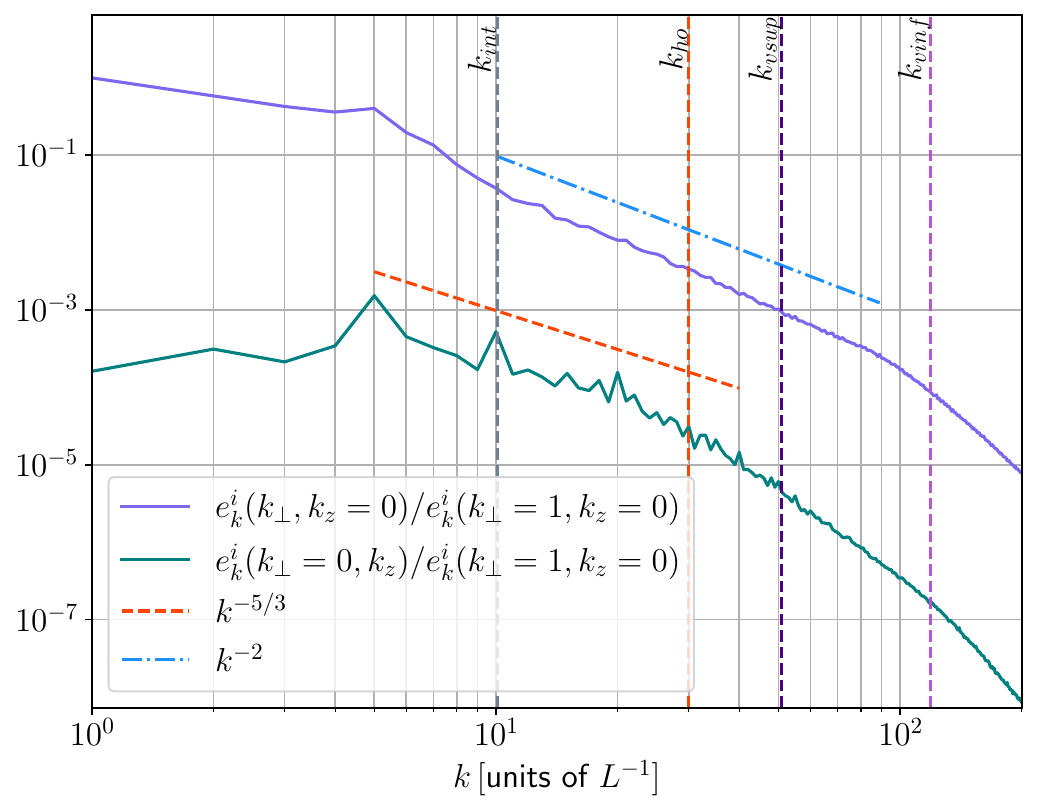}
    \caption{Incompressible kinetic energy spectrum (for $\Omega=1.2$) as a function of $k_\perp$ for modes with $k_z=0$ (top spectrum), and as a function of $k_z$ for modes with $k_\perp=0$ (bottom spectrum), normalized by the incompressible energy in $(k_\perp = 1, k_z=0)$. Power laws and characteristic wave numbers are indicated as references.} 
    \label{fig: spectrums kz kperp }
\end{figure}

Anisotropic sound (or compressible mode) emission was observed in experiments of non-turbulent rotating BECs \cite{Simula2005}. In our case they seem to provide different mechanisms for the energy dissipation, along different direcions in spectral space. Figure \ref{fig: spectrums kz kperp } shows the incompressible kinetic spectrum $e_k^\textrm{(i)}(k_\perp, k_z)$, for modes with $k_\perp=0$ or $k_z=0$, and for $\Omega=1.2$. These spectra can be computed from the full spatio-temporal spectrum $E_k^\textrm{(i)}({\bf k}, \omega)$ by integrating over all frequencies. In the direction of $k_\perp$ the spectrum displays a scaling compatible with $\sim k_\perp^{-2}$ in a broad range of wave numbers (compatible with predictions from the theory of classical rotating turbulence \cite{BELLET2006}), while along $k_z$ the spectrum displays a $\sim k_z^{-5/3}$ compatible scaling. This scaling is visible at wave numbers above and below the intervortex wave number, and thus is probably the result of vortex reconnection with some contribution of a Kelvin wave cascade.

\section{Conclusions}
\label{sec: conclusion}

Rotating quantum turbulence is fundamentally different from both non-rotating quantum turbulence, as well as from classical rotating turbulence. The quasi-two-dimensionalization of the flow results in an inverse energy transfer, as in quasi-2D quantum turbulence \cite{Mller2020} and in classical rotating turbulence \cite{Sen_2012}. This inverse transfer can be also interpreted as a negative temperature state, as predicted for 2D point vortices \cite{Onsager1949} and observed in BEC experiments \cite{Gauthier_2019, Johnstone_2019}. However, the small scales display a scaling different from all other regimes.

A $\sim k^{-1}$ power law at intermediate wave numbers in the incompressible kinetic energy is reminiscent of the scaling of Vinen turbulence, albeit in this case there is no obvious counterflow in the system. However, the system displays very little transfer of energy to small scales (most kinetic energy is transferred to larger scales), and a different arrangement of quantized vortices. This, together with a pulsation of the condensate inwards and outwards (with the associated friction of the vortices with this flow), can provide a way for the system to dissipate energy in the perpendicular direction as suggested by the spatio-temporal spectra. Along the axis of rotation, energy is dissipated instead as sound waves. This results in a thermalization of one-dimensional sound modes, with a flat spectrum of the compressible kinetic energy, and distinct scaling of the incompressible energy when individual modes are studied: a $\sim k_z^{-5/3}$ subdominant scaling for modes with $k_\perp=0$, and a $\sim k_\perp^{-2}$ dominant scaling for modes with $k_z=0$. A similar mechanism may be also present in recent simulations of quantum turbulence in BECs \cite{Cidrim2017, Marino2021}, in which cigar-shaped traps and a few multicharged aligned vortices are studied.

\begin{acknowledgments}
JAE and PDM acknowledge financial support from UBACYT Grant No.~20020170100508BA and ANPCyT PICT Grant No.~2018-4298. MB acknowledges support from the French Agence Nationale de la Recherche (ANR QUTE-HPC project No.~ANR-18-CE46-0013).
\end{acknowledgments}

\bibliography{ms}

\end{document}